\documentclass[10pt]{article}

\usepackage{amsmath}
\usepackage{amssymb}

\usepackage{graphicx}
\usepackage[hang,tight]{subfigure}

\usepackage{cite,url}

\usepackage{color}
\usepackage[labelfont=bf,labelsep=period,justification=raggedright]{caption}

\makeatletter
\renewcommand{\@biblabel}[1]{\quad#1.}
\makeatother

\date{}

\begin{document}
\vspace*{0.35in}

% Please capitalize all terms in the title except conjunctions, prepositions, and articles.
\begin{flushleft}
{\Large
\textbf{Statistical Properties and Pre-hit Dynamics of Price Limit Hits in the Chinese Stock Markets}
}
\newline
\\
%Name1 Surname\textsuperscript{1,\Yinyang},
%Name2 Surname\textsuperscript{2,\Yinyang},
%Name3 Surname\textsuperscript{2,\textcurrency a},
%Name4 Surname\textsuperscript{2,\ddag},
%Name5 Surname\textsuperscript{2,\ddag},
%Name6 Surname\textsuperscript{2},
%Name7 Surname\textsuperscript{3,*}

Yu-Lei Wan\textsuperscript{1,2},
Wen-Jie Xie\textsuperscript{2,3,4},
Gao-Feng Gu\textsuperscript{2,3},
Zhi-Qiang Jiang\textsuperscript{2,3},
Wei Chen\textsuperscript{5},
Xiong Xiong\textsuperscript{6,7,*},
Wei Zhang\textsuperscript{6,7},
Wei-Xing Zhou\textsuperscript{1,2,3,*}

\bigskip

\bf{1} Department of Mathematics, School of Science, East China University of Science and Technology, Shanghai 200237, China
\\
\bf{2} Research Center for Econophysics, East China University of Science and Technology, Shanghai 200237, China
\\
\bf{3} Department of Finance, School of Business, East China University of Science and Technology, Shanghai 200237, China
\\
\bf{4} Postdoctoral Research Station, School of Social and Public Administration, East China University of Science and Technology, Shanghai 200237, China
\\
\bf{5} Shenzhen Stock Exchange, Shenzhen 518010, China
\\
\bf{6} Collage of Management and Economics, Tianjin University, Tianjin 300072, China
\\
\bf{7} China Center for Social Computing and Analytics, Tianjin University, Tianjin 300072, China
\\
%$\ast$ E-mail: xxpeter@tju.edu.cn (XX); wxzhou@ecust.edu.cn (WXZ)
%\\
\bigskip

% Insert additional author notes using the symbols described below. Insert symbol callouts after author names as necessary.
%
% Remove or comment out the author notes below if they aren't used.
%
% Primary Equal Contribution Note
%\Yinyang These authors contributed equally to this work.

% Additional Equal Contribution Note
% Also use this double-dagger symbol for special authorship notes, such as senior authorship.
%\ddag These authors also contributed equally to this work.

% \textcurrency a Insert current address of first author with an address update
% \textcurrency b Insert current address of second author with an address update
% \textcurrency c Insert current address of third author with an address update

% Group/Consortium Author Note
%\textpilcrow Membership list can be found in the Acknowledgments section.

% Use the asterisk to denote corresponding authorship and provide email address in note below.
* xxpeter@tju.edu.cn (XX); wxzhou@ecust.edu.cn (WXZ)

\end{flushleft}
\section*{Abstract}

Price limit trading rules are adopted in some stock markets (especially emerging markets) trying to cool off traders' short-term trading mania on individual stocks and increase market efficiency. Under such a microstructure, stocks may hit their up-limits and down-limits from time to time. However, the behaviors of price limit hits are not well studied partially due to the fact that main stock markets such as the US markets and most European markets do not set price limits. Here, we perform detailed analyses of the high-frequency data of all A-share common stocks traded on the Shanghai Stock Exchange and the Shenzhen Stock Exchange from 2000 to 2011 to investigate the statistical properties of price limit hits and the dynamical evolution of several important financial variables before stock price hits its limits. We compare the properties of up-limit hits and down-limit hits. We also divide the whole period into three bullish periods and three bearish periods to unveil possible differences during bullish and bearish market states. To uncover the impacts of stock capitalization on price limit hits, we partition all stocks into six portfolios according to their capitalizations on different trading days. We find that the price limit trading rule has a cooling-off effect (object to the magnet effect), indicating that the rule takes effect in the Chinese stock markets. We find that price continuation is much more likely to occur than price reversal on the next trading day after a limit-hitting day, especially for down-limit hits, which has potential practical values for market practitioners.

% Please keep the Author Summary between 150 and 200 words
% Use first person. PLOS ONE authors please skip this step.
% Author Summary not valid for PLOS ONE submissions.
%\section*{Author Summary}
%Lorem ipsum dolor sit amet, consectetur adipiscing elit. Curabitur eget porta erat. Morbi consectetur est vel gravida pretium. Suspendisse ut dui eu ante cursus gravida non sed sem. Nullam sapien tellus, commodo id velit id, eleifend volutpat quam. Phasellus mauris velit, dapibus finibus elementum vel, pulvinar non tellus. Nunc pellentesque pretium diam, quis maximus dolor faucibus id. Nunc convallis sodales ante, ut ullamcorper est egestas vitae. Nam sit amet enim ultrices, ultrices elit pulvinar, volutpat risus.

%\linenumbers

\section*{Introduction}

In many stock markets, price limit rules are set expecting to reduce remarked swings by cooling off traders' irrational mania. A stable stock market has lower risks and thus attracts more people to participate. This is certainly increase the resource reallocation function of stock markets and benefits the economies. Price limit rules constrain intraday prices to move within a preset price interval embraced by a price up limit and a price down limit. Usually, the limit prices are determined by fixed fluctuation percentages in reference to the closing price of the previous day. In most stock markets, the fluctuation percentages for up-limit and down-limit are symmetric. However, there are also examples for asymmetric price limits especially in certain market states. After the price reaches its limit, a circuit breaker may be triggered causing trading halt in some markets, while in other markets the traders can continue to trade shares \cite{Subrahmanyam-1994-JF}.

The effectiveness of the price limit rules is controversial. It is expected to have a cooling-off effect to reduce the volatility of stocks \cite{Ma-Rao-Sears-1989-JFSR}. On the contrary, it may also cause a magnet effect, which refers to the phenomenon that the price limit acts as a magnet to attract more trades leading to higher trading intensity and price volatility and increases the probability of price rise or fall when the price is closer to the limit price \cite{Subrahmanyam-1994-JF}. The magnet effect occurs when the traders fear of the lack of liquidity and possible position lock caused by imminent price limit hits, and the traders are thus eager to protect themselves through submitting aggressive sub-optimal orders, which usually induces large price variations and heavy trading volumes. Since the cooling effect or the magnet effect takes place at the intraday level, studies of the presence of either effect, the evolution of pre-hit dynamics and the performance of price limit rules are of great interest to academics, investors and regulators to gain a better understanding of the mechanisms of how the market structure and the investors' trading behavior affects price discovery.

The study of price limits started on futures markets \cite{Telser-1981-JFutM,Brennan-1986-JFE}. Empirical analysis has been carried out for different markets at different time periods. There is no consensus on the presence of a magnet effect or a cooling-off effect. Arak and Cook investigated if price behavior is infected by price limits on the treasury bond futures market and found no evidence of a magnet effect but rather a reversal effect \cite{Arak-Cook-1997-JFSR}. Berkman and Steenbeek compared the price formation processes under different price limits between Osaka Securities Exchange and Nikkei 225 index on the Singapore International Monetary Exchange and found no significant arbitrage opportunities between the two markets \cite{Berkman-Steenbeek-1998-JFinM}. In recent years, empirical studies about the magnet effect concentrated on stock markets. Cho et al. studied the 5-min return time series of 345 stocks traded on the Taiwan Stock Exchange from 1998/01/03 to 1999/03/20 and reported a statistically and economically significant magnet effect for stock prices to accelerate towards the up-limit and weak evidence of acceleration towards the down-limit \cite{Cho-Russell-Tiao-Tsay-2003-JEF}. Hsieh et al. analyzed the transaction data of 439 stocks traded on the Taiwan Stock Exchange in 2000 using logit models and found evidence of the magnet effect on both up-limt and down-limit \cite{Hsieh-Kim-Yang-2009-JEF}. Du et al. also observed the magnet effect when the stock prices are approaching the price limits in the Korean market \cite{Du-Liu-Rhee-2005-WP}.

The Chinese stock markets also set price limits, which varied over time. The current $\pm10\%$ price limits were fixed since 1996/12/16 for A-share common stocks. There are several studies conducted on the presence of the magnet or cooling-off effect. However, empirical results lead to controversial conclusions \cite{Li-2005-cnJCUFE,Meng-Jiang-2008-cnTE,Fang-Chen-2007-cnSE,Wong-Liu-Zeng-2009-CER,Zeng-She-2014-cnASM,Zhang-Zhu-2014-cnJCQUT}. According to the data released by the China Securities Regulatory Commission, by June 2014, there are 2540 listed companies and the total market capitalization is about 24.412 trillion Chinese Yuan. Due to its huge capitalization and representativeness as an emerging market, research on Chinese stocks is of great importance and remarkable interest. In this work, we will perform detailed analyses on the statistical properties of variables associated with price limit hits and the pre-hit dynamics of important financial variables before limit hits in the Chinese stock markets. These issues are less studied in previous works. To obtain conclusive results about the presence of a magnet or cooling-off effect in the Chinese stock markets, one needs to adopt different methods proposed in the literature and consider possible evolution of the effect (if present) per se. We leave this topic in a future work.

% You may title this section "Methods" or "Models".
% "Models" is not a valid title for PLoS ONE authors. However, PLoS ONE
% authors may use "Analysis"
\section*{Materials and Methods}

\subsection*{The Chinese stock markets}

There are two stock exchanges in mainland China. The first market for government approved securities was founded in Shanghai on 1990/11/26 and started operation on 1990/12/19 under the name of the Shanghai Stock Exchange (SHSE). Shortly after, the Shenzhen Stock Exchange (SZSE) was established on 1990/12/01, and started its operations on 1991/07/03. There are two separate markets for A-shares and B-shares on both exchanges. A-shares are common stocks issued by mainland Chinese companies, subscribed and traded in Chinese currency Renminbi (RMB), listed on mainland Chinese stock exchanges, bought and sold by Chinese nationals and approved foreign investors. B-shares are issued by mainland Chinese companies, traded in foreign currencies and listed on mainland Chinese stock exchanges. B-shares carry a face value denominated in RMB. The B-share market was launched in 1992 and was restricted to foreign investors before 2001/02/19. It has been open to Chinese investors since. The microstructure of the two markets has been changed on several aspects, such as the daily price up/down limit rules imposed since 1996/12/16. The price limits are $\pm10\%$ for common stocks and $\pm5\%$ for specially treated (ST and ST*) stocks. We note that, before 1996/12/16, there were also periods with different intervals of price limits or without price limits.

On each trading day, the trading time period is divided into three parts: opening call auction, cooling periods, and continuous double auction. The market opens at 9:15 a.m. and enters the opening call auction until 9:25 a.m, during which the trading system accepts order submissions and cancelations, and all matched transactions are executed at 9:25 a.m. This is followed by a cooling period from 9:25 to 9:30 a.m. During the cooling period, the exchanges are open to order routing from members, but does not accept the cancelation of orders. All matched orders are executed in real time. However, the information is not released to trading terminals during the cooling period and is publicly available at the end of the cooling period. The continuous double auction operates from 9:30 to 11:30 and from 13:00 to 15:00 (for SZSE, 14:57-15:00 is a closing call auction period to form the close price) and transaction occurs automatically by matching due to price and time priority. The time interval between 11:30 a.m. and 13:00 p.m. is a trade halt period. Outside these opening hours, unexecuted orders will be removed by the system.

\subsection*{Data sets}

Our data sets were provided by RESSET (http://resset.cn/), which is a leading financial data provider supporting academic research. The data sets contain all common A-share stocks traded on the Shanghai Stock Exchange and Shenzhen Stock Exchange. The price limits for these stocks are $\pm10\%$. Specially treated stocks with price limits of $\pm5\%$ are not included in our analysis. The sample covers the period from 2000/01/04 to 2011/12/30, totally 12 years. Because the stocks have different initial public offering dates, the lengthes of stocks are not fixed. The quote frequency is about 5 seconds before 27 June 2011 and 3 seconds afterwards. Due to different liquidities of the stocks, the quote frequencies of different stocks can be lower. For each stock, we have a unique stock code that is a sequence of six digital numbers, the trading time, the trading price, the trading volume, and the prices and standing volumes at the three best levels before 5 December 2003 or 5 levels afterwards on both the buy and sell sides of the limit order book.

\subsection*{Determining daily price up/down limits}

The records of stocks do not contain any indicator of price hits. Hence, we need to identify when the price of a stock hits the up-limit or the low limit. Because the tick size of all stocks is one cent (0.01 Chinese Yuan), no matter they have high prices or low prices, we are able to identify price limit hits. Indeed, for each stock, we can determine the price up limit and price down limit for each trading day. Let $P_i(T)$ denotes the closing price of stock $i$ on day $T$. The up-limit $P^+_i(T+1)$ and the down-limit $P^-_i(T+1)$ of stock $i$ on day $T+1$ are determined as follows,
\begin{equation}
  P^{\pm}_i(T+1) = {\mathcal{R}\left[100P_i(T)(1\pm10\%)\right]}/100,
  \label{Eq:P+-}
\end{equation}
where $\mathcal{R}\left[x\right]$ is a round operator of $x$ such that the daily price limits are rounded to the tick size according to the {\textit{Trading Rules of Shanghai Stock Exchange}} (2003, 2006) and the {\textit{Shenzhen Stock Exchange Trading Rules}} (2003, 2006).

%
%For each stock, we identify the trading days that have price limit hits. And we define price limit day as follows,
%\begin{equation}
% \label{Eq:Price:Hit:definition}
%\left\{ \begin{array}{l}
%\frac{{Tp_{H,i} + 0.005}}{{Preclpr}} > 1.1,~~~~limit-up~~events\\
%\frac{{Tp_{L,i} - 0.005}}{{Preclpr}} < 0.9,~~~~limit-down~~ events
%\end{array} \right.
%\end{equation}
%Where $Tp_{H,i}$,$Tp_{L,i}$ means the highest and lowest price on trading day $i$, $Preclpr$ means the previous session's closing price before trading day $i$.

\subsection*{Determining bullish and bearish periods of the Chinese stock markets}

\begin{figure}[htb]
  \centering
  \includegraphics[width=8cm,height=6cm]{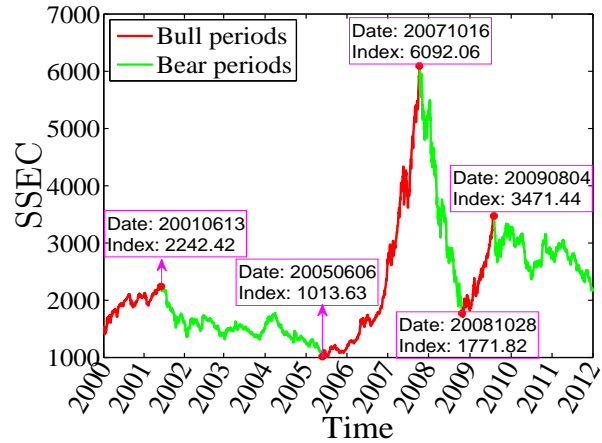}
  \caption{\label{Fig:SSEC:Bull:Bear} {\textbf{Evolution of the Shanghai Stock Exchange Composite Index from January 2000 to December 2011.}} The data shown are the closing prices. The historical highs and lows usually occurred intraday. The index has been divided into alternating bullish and bearish periods. The red parts stand for bullish periods and the green parts correspond to bearish periods.}
\end{figure}

A representative measure of the status of the Chinese stock markets is the Shanghai Stock Exchange Composite (SSEC) Index. Figure \ref{Fig:SSEC:Bull:Bear} illustrates the evolution of the SSEC Index from 1999 to 2011. The SSEC Index rose from 1406 on 2000/01/04 to its historical intraday high of 2242.4 on 2001/06/13 and since plummeted 32.2\% to 1520.7 on 2001/10/22. The Chinese stock markets entered a bearish antibubble state since June of 2011 \cite{Zhou-Sornette-2004a-PA}, which did not synchronize the U.S. antibubble after the burst of the so-called New Economy Bubble in 2000 \cite{Johansen-Sornette-2000a-EPJB,Sornette-Zhou-2002-QF}. The SSEC Index hit its all-time intraday low at 998.23 on 2005/06/04 during the time period investigated in this work. After that, A huge bubble formed and the index reached its historical intraday high at 6124.04 on 2007/10/16, which was followed by a severe crash \cite{Jiang-Zhou-Sornette-Woodard-Bastiaensen-Cauwels-2010-JEBO} and the index plummeted to 1666.93 on 2008/10/28 as a all-time low since the crash. Although the whole individual sentiment was bearish in the last six years, there was a mediate-size bubble started after 2008/10/28 and the index reached 3478.01 on 2009/08/04. In summary, the time period under investigation in this work can be divided into alternating periods of bullish and bearish states. The stock market was bullish during the three time periods: 2000/1/4 - 2001/6/13, 2005/06/04 - 2007/10/16, and 2008/10/28 - 2009/08/04. Other periods are recognized as being bearish.

In this work, we will compare the statistics of price limit hits in bullish and bearish market states. It is true that individual stocks may evolve differently. However, it is hard to recognize bullish and bearish states for individual stocks and it is not irrational to make such a comparison based on the SSEC Index. We leave this more complicated stock-by-stock classification of bulls and bears in future research.

\subsection*{Fitting procedure}

The empirical distribution of the number of limit hits for individuals is truncated on the left. We find that most of the distributions investigated in this work can be fitted by the left truncated normal distribution:
\begin{equation}
  p(x) = \frac{1}{\sqrt{2\pi}\sigma}e^{\frac{(x-\mu)^2}{2\sigma^2}}, ~~~x>0.
  \label{Eq:Normal:Left}
\end{equation}
Both ordinary least squares and maximum likelihood estimation are applied in curve fitting. When we use maximum likelihood estimation to estimate parameters such as mean and variance, we cannot use the sample mean and variance to substitute the whole population. A feasible method is the following \cite{Randall-J.Olsen-1980-Em}:
\begin{equation}
  \label{Eq:Ex:Qr}
  \left\{ \begin{array}{l}
  E(X|X > 0) = \mu  + \sigma Q(\mu /\sigma )\\
  Q(r) = {{f(r)}}/{{F(r)}}
\end{array} \right.\
\end{equation}
where $f(\cdot)$ and $F(\cdot)$ are respectively the density function and the cumulative density function of the standard normal distribution. Denoting $\mu' = E(X|X > 0)$ the mean of the truncated normal distribution, one has
\begin{equation}
\label{Eq:mu:var}
\left\{ \begin{array}{l}
  \mu' = \mu  + \sigma Q(\mu /\sigma )\\
  {\mathrm{Var}}(X|X>0) = {\sigma^{2}}\{1-Q(\mu /\sigma )[\mu /\sigma + Q(\mu /\sigma )]\}
\end{array} \right.\
\end{equation}
Denoting ${\sigma'^{2}} = {\mathrm{Var}}(X|X > 0)$ the variance of the truncated normal distribution, one obtains
\begin{equation}
\label{Eq:Mu:Sigma}
 \mu'/\sigma'= \frac{\mu/\sigma + Q(\mu/\sigma)}{\sqrt{\{1-Q(\mu /\sigma )[\mu /\sigma + Q(\mu /\sigma )]\}}}.
\end{equation}
Defining $r=\mu/\sigma$, $r'=\mu'/\sigma'$, one gets
\begin{equation}
\label{Eq:R:result}
 r'= \left[r+ \frac{{f(r)}}{{F(r)}} \right]\div\sqrt{1-\frac{{f(r)}}{{F(r)}}\left[r + \frac{{f(r)}}{{F(r)}}\right]}.
\end{equation}

In practical applications, we find that the function including $r$ on the right side of Eq.~(\ref{Eq:R:result}) is monotonically increasing. Hence the equation has a unique solution. We can determine $r$ firstly, and then $\sigma$ and $\mu$. Finally, we use $\sigma$ and $\mu$ to estimate the population distribution.

% Results and Discussion can be combined.
\section*{Results}

\subsection*{Basic statistics of the numbers of limit-hitting days}

We provide statistical properties of the numbers of trading days with different types of limit hits. The variables are the following. $N$ is the total number of trading days with limit hits. $\langle{N}\rangle$ is the average number of limit-hitting days for individual stocks. $N_{\rm{con}}$ is the number of limit-hitting days with continued next-day opening prices. It contains up (down) limit hitting days with the opening prices on next trading days being higher (lower) than the closing prices. $N_{\rm{rev}}$ is the number of limit-hitting days with price reversal on the next day. It contains up (down) limit hitting days with the opening prices on next trading days being lower (larger) than the closing prices. $N_{\rm{open}}$, $N_{\rm{am}}$, $N_{\rm{pm}}$ and $N_{\rm{close}}$ are respectively the numbers of days with limit hits occurred in the opening call auction (9:15, 9:30], in the continuous double auction session (9:30,11:30] in the morning, in the continuous double auction session [13:00, 15:00] in the afternoon and at the closure of the trading days. $N_{\rm{close,con}}$ and $N_{\rm{close,rev}}$ are the numbers of trading days that closed at limit prices and the price continued rising up or falling down on the successive trading days. It is possible that for some stocks there are both up- and down-limit hits within the same day. In this case, the first limit hit is used in the calculation of different numbers. We further delete IPO days and ex-dividend days. Since trading halt may trigger after a price limit-hitting day and thus there is no followup open price, we do not count these days in $N_{\rm{con}}$, $N_{\rm{rev}}$, $N_{\rm{close,con}}$, and $N_{\rm{close,rev}}$. We also partition evenly all stocks with limit hits on a given day into six portfolios based on their capitalizations (Portfolio 1 with the smallest capitalizations and Portfolio 6 with the largest capitalizations) and count the defined numbers for each portfolio. The numbers are determined for the whole period, and the bullish and bearish periods as well. The capitalization of a stock is calculated as the product of the amount of shares times the price. The basic statistics of limit hits in the whole sample period and in bullish and bearish periods are presented in Table~\ref{Tb:Statistics:ManyStocks}. It is trivial to observe that
\begin{equation}
  N_{\pm} = \sum_{j=1}^6 N_{\pm,j}~~{\mathrm{and}}~~N_{\pm,i}\approx N_{\pm,j},
\end{equation}
where the subscripts $+$ and $-$ represent respectively price up limit hit and price down limit hit, and $j$ stands for the portfolio serial number.

\setlength\tabcolsep{2.5pt}
\begin{table}[!ht]
%  \footnotesize
%  \begin{adjustwidth}{-2.25in}{0in} % Comment out/remove adjustwidth environment if table fits in text column.
  \caption{{\textbf{Summary statistics of limit hits in the whole period (Panel A), in the bullish periods (Panel B), and in the bearish periods (Panel C).}} $N$ is the total number of trading days with limit hits. $\langle{N}\rangle$ is the average number of limit-hitting days for individual stocks. $N_{\rm{con}}$ is the number of limit-hitting days with continued next-day opening prices (up-limit days with the next-day opening prices higher than the closing prices and down-limit days with the next-day opening price lower than the closing prices). $N_{\rm{rev}}$ is the number of limit-hitting days with price reversal on the next day. $N_{\rm{open}}$, $N_{\rm{am}}$, $N_{\rm{pm}}$ and $N_{\rm{close}}$ are respectively the numbers of days with limit hits occurred in the opening call auction, in the continuous double auction in the morning, in the continuous double auction in the afternoon and at the closure of the trading days. $N_{\rm{close,con}}$ and $N_{\rm{close,rev}}$ are the numbers of trading days that closed at limit prices and the price continued rising or falling on the successive trading days. }
  \begin{tabular}{cccccccccccccccccccccccccccccc}
\hline\hline
\multicolumn{8}{l}{{\small{Panel A: Whole sample period}}}\\\hline
  && \multicolumn{2}{c}{All stocks} && \multicolumn{2}{c}{Portfolio 1} && \multicolumn{2}{c}{Portfolio 2} && \multicolumn{2}{c}{Portfolio 3} && \multicolumn{2}{c}{Portfolio 4} && \multicolumn{2}{c}{Portfolio 5} && \multicolumn{2}{c}{Portfolio 6}\\
  \cline{3-4}\cline{6-7}\cline{9-10}\cline{12-13}\cline{15-16}\cline{18-19}\cline{21-22}
  &&Up & Down && Up & Down && Up & Down && Up & Down && Up & Down && Up & Down  && Up & Down  \\\hline
  $N$                      && 62346& 44216&& 10393& 7369&& 10393& 7369&& 10390& 7368&& 10389& 7368&& 10387& 7368&& 10394&7374\\
  $\langle{N}\rangle$      && 20.58 & 14.60 && 3.43 & 2.43 && 3.43 & 2.43 && 3.43 & 2.43 && 3.43 & 2.43 && 3.43 & 2.43 &&3.43 & 2.43\\
  $N_{\rm{con}}$           && 39696& 33601&& 6595& 6242&& 6819& 5706&& 6662& 5569&& 6538& 5514&& 6443& 5324&& 6639& 5246\\
  $N_{\rm{rev}}$           &&22627& 10603&& 3798& 1127&& 3574& 1662&& 3724& 1798&& 3848& 1852&& 3937& 2040&& 3746& 2124\\
  $N_{\rm{open}}$            &&6718& 2126&& 1444& 441&& 1445& 332&& 1115& 334&& 898& 347&& 886& 316&& 930& 356\\
  $N_{\rm{am}}$            &&33204& 15694&& 5740& 3082&& 5831& 2460&& 5673& 2551&& 5522& 2674&& 5434& 2496&& 5004& 2431\\
  $N_{\rm{pm}}$            &&29157& 28525&& 4653& 4287&& 4562& 4909&& 4720& 4818&& 4871& 4695&& 4959& 4873&& 5392& 4943\\
  $N_{\rm{close}}$         &&40752& 22213&& 6207& 3203&& 6948& 3745&& 7137& 3882&& 7004& 3857&& 6871& 3791&& 6585& 3735\\
  $N_{\rm{close,con}}$     &&  32340& 19067&& 5120& 2986&& 5610& 3252&& 5612& 3316&& 5480& 3278&& 5321& 3147&& 5197&3088\\
  $N_{\rm{close,rev}}$     &&8394& 3138&& 1087& 217&& 1338& 493&& 1521& 566&& 1522& 577&& 1544& 641&& 1382& 644\\\hline
  \hline\multicolumn{8}{l}{Panel B: Bull period}\\\hline
  && \multicolumn{2}{c}{All stocks} && \multicolumn{2}{c}{Portfolio 1} && \multicolumn{2}{c}{Portfolio 2} && \multicolumn{2}{c}{Portfolio 3} && \multicolumn{2}{c}{Portfolio 4} && \multicolumn{2}{c}{Portfolio 5} && \multicolumn{2}{c}{Portfolio 6}\\
  \cline{3-4}\cline{6-7}\cline{9-10}\cline{12-13}\cline{15-16}\cline{18-19}\cline{21-22}
  &&Up & Down && Up & Down && Up & Down && Up & Down && Up & Down && Up & Down  && Up & Down  \\\hline
  $N$                      &&32593& 17898&& 5432& 2983&& 5432& 2983&& 5432& 2983&& 5432& 2983&& 5431& 2983&& 5434& 2983\\
  $\langle{N}\rangle$      && 26.52 & 14.56 && 4.42 & 2.43 && 4.42 & 2.43 && 4.42 & 2.43 && 4.42 & 2.43 && 4.42 & 2.43 && 4.42 & 2.43\\
  $N_{\rm{con}}$           &&22344& 13888&& 3501& 2563&& 3713& 2397&& 3690& 2315&& 3787& 2278&& 3775& 2190&& 3878& 2145\\
  $N_{\rm{rev}}$           && 10249& 4010&& 1931& 420&& 1719& 586&& 1742& 668&& 1645& 705&& 1656& 793&& 1556& 838\\
 $N_{\rm{open}}$            && 2908& 986&& 660& 161&& 508& 134&& 421& 175&& 432& 183&& 440& 162&& 447& 171\\
  $N_{\rm{am}}$            &&15785& 6730&& 2925& 1117&& 2657& 1005&& 2650& 1168&& 2626& 1191&& 2590& 1173&& 2337& 1076\\
  $N_{\rm{pm}}$            && 16809& 11168&& 2507& 1866&& 2775& 1978&& 2782& 1815&& 2806& 1792&& 2842& 1810&& 3097& 1907\\
  $N_{\rm{close}}$         &&20814& 8778&& 3035& 1278&& 3511& 1480&& 3644& 1559&& 3652& 1540&& 3577& 1493&& 3395& 1428\\
  $N_{\rm{close,con}}$     &&17394& 8011&& 2541& 1195&& 2923& 1371&& 2977& 1407&& 3062& 1406&& 2986& 1352&& 2905& 1280\\
  $N_{\rm{close,rev}}$     &&3420& 767&& 494& 83&& 588& 109&& 667& 152&& 590& 134&& 591& 141&& 490& 148\\
  \hline\multicolumn{8}{l}{Panel C: Bear period}\\\hline
  && \multicolumn{2}{c}{All stocks} && \multicolumn{2}{c}{Portfolio 1} && \multicolumn{2}{c}{Portfolio 2} && \multicolumn{2}{c}{Portfolio 3} && \multicolumn{2}{c}{Portfolio 4} && \multicolumn{2}{c}{Portfolio 5} && \multicolumn{2}{c}{Portfolio 6}\\
  \cline{3-4}\cline{6-7}\cline{9-10}\cline{12-13}\cline{15-16}\cline{18-19}\cline{21-22}
  &&Up & Down && Up & Down && Up & Down && Up & Down && Up & Down && Up & Down  && Up & Down  \\\hline
  $N$                      && 29753& 26318&& 4961& 4386&& 4960& 4386&& 4956& 4385&& 4957& 4385&& 4958& 4385&& 4961& 4391\\
  $\langle{N}\rangle$      && 16.53 & 14.62 && 2.76 & 2.44 && 2.76 & 2.44 && 2.75 & 2.44 && 2.75 & 2.44 && 2.75 & 2.44 && 2.76 & 2.44\\
  $N_{\rm{con}}$           &&17352& 19713&& 3062& 3678&& 3034& 3299&& 2810& 3247&& 2785& 3231&& 2824& 3162&& 2837& 3096\\
  $N_{\rm{rev}}$           && 12378& 6593&& 1899& 708&& 1926& 1086&& 2141& 1137&& 2167& 1152&& 2130& 1219&& 2115& 1291\\
  $N_{\rm{open}}$            &&3810& 1140&& 935& 282&& 931& 189&& 600& 155&& 437& 165&& 409& 163&& 498& 186\\
  $N_{\rm{am}}$            && 17419& 8964&& 3004& 1960&& 3231& 1434&& 2984& 1404&& 2843& 1445&& 2747& 1365&& 2610& 1356\\
  $N_{\rm{pm}}$            && 12348& 17357&& 1957& 2426&& 1730& 2952&& 1977& 2982&& 2118& 2941&& 2214& 3021&& 2352& 3035\\
  $N_{\rm{close}}$         && 19938& 13435&& 3244& 1924&& 3424& 2261&& 3432& 2324&& 3344& 2322&& 3333& 2311&& 3161& 2293\\
  $N_{\rm{close,con}}$     &&14946& 11056&& 2621& 1788&& 2644& 1871&& 2502& 1900&& 2423& 1857&& 2435& 1824&& 2321& 1816\\
  $N_{\rm{close,rev}}$     &&4974& 2371&& 623& 136&& 780& 390&& 925& 424&& 917& 463&& 895& 484&& 834& 474\\
  \hline\hline
  \end{tabular}
%  \begin{flushleft} Table notes Phasellus venenatis, tortor nec vestibulum mattis, massa tortor interdum felis, nec pellentesque metus tortor nec nisl. Ut ornare mauris tellus, vel dapibus arcu suscipit sed.
%  \end{flushleft}
  \label{Tb:Statistics:ManyStocks}
%  \end{adjustwidth}
\end{table}

Panel A of Table \ref{Tb:Statistics:ManyStocks} shows the results for the whole period from 2000 to 2011. There are more limit-up hits than limit-down hits, which is partially caused by the rapid growth of China's economy and the absence of short mechanism. The limit-down days are more likely to have next-day price continuation than limit-up days since $N^+_{\rm{con}}/N^+=39696/62346=63.7\%$ and $N^-_{\rm{con}}/N^-=33601/44216=75.9\%$. In addition, a limit hitting day is more likely to have price continuation than price reversal because $N^{\pm}_{\rm{con}}\gg N^{\pm}_{\rm{rev}}$. For all other numbers concerning all stocks and different portfolios, we also observe more limit-up days than limit-down days. For limit-down days, the probability of next-day price continuation decreases with increasing average capitalization of the portfolio, while the probability of next-day price reversal increases with the capitalization. For limit-up days, the probabilities of next-day price continuation and price reversal do no have any clear trend. For $N^{\pm}_{\rm{open}}$ and $N^{\pm}_{\rm{am}}$, we observe decreasing trends with capitalization. For $N^{\pm}_{\rm{pm}}$, we observe increasing trends with capitalization. For $N^{\pm}_{\rm{close}}$, they increase first and then decrease. In all the seven cases (all stocks and the six portfolios), limit-up events are more likely to occur in the morning than in the afternoon, while limit-down events are more likely to happen in the afternoon than in the morning. Limit-up events have higher probability (65.3\%) to close at limit price than limit-down events (51.0\%) according to $N_{\rm{close}}$. When a trading day closes at the up-limit or the down-limit, the next-day opening price will rise with a very high probability (79.4\% for limit-up events and 85.8\% for limit-down events). In addition, $N^{\pm}_{\rm{close,con}}$ and $N^{\pm}_{\rm{close,rev}}$ increase first and then decrease.

For the bullish periods in Panel B of Table \ref{Tb:Statistics:ManyStocks}, in all seven cases (all stocks and the six portfolios), there are more limit-up days for $N$, $N_{\rm{open}}$, $N_{\rm{am}}$ and $N_{\rm{close}}$. For the bullish periods in Panel C of Table \ref{Tb:Statistics:ManyStocks}, there are more limit-up days than limit-down days for $N$, $N_{\rm{open}}$, $N_{\rm{am}}$, and $N_{\rm{close}}$, with an exception that there are more limit-down days than limit-up days in the afternoon ($N^-_{\rm{pm}}>N^+_{\rm{pm}}$). However, the occurrence difference between limit-up and limit-down days is smaller in the bearish periods than in the bullish periods. For bullish periods in Panel B, $N^+_{\rm{open}}$ and $N^+_{\rm{am}}$ decrease with increasing capitalization, $N^+_{\rm{pm}}$ increases with capitalization, and no evident trends have been found in $N^-_{\rm{open}}$, $N^-_{\rm{am}}$, $N^-_{\rm{pm}}$ and $N^{\pm}_{\rm{close}}$. For bearish periods in Panel C, $N^{\pm}_{\rm{open}}$ and $N^{\pm}_{\rm{am}}$ decrease with capitalization, $N^+_{\rm{pm}}$ and $N^-_{\rm{pm}}$ increase with capitalization, and $N^+_{\rm{close}}$ and $N^{-}_{\rm{close}}$ do not exhibit any clear trend with respect to capitalization.

For both bullish and bearish periods, price continuation is more likely to occur than price reversal for both limit-up and limit-down days, because $N^{\pm}_{\rm{con}}>N^{\pm}_{\rm{rev}}$ and $N^{\pm}_{\rm{close,con}}>N^{\pm}_{\rm{close,rev}}$. In addition, the probability of price continuation is higher for limit-down days than for limit-up days, that is $N^{+}_{\rm{con}}/N^+<N^-_{\rm{con}}/N^-$ and $N^{+}_{\rm{close,con}}/N^+_{\rm{close}}<N^-_{\rm{close,con}}/N^-_{\rm{close}}$. For bullish periods, the probability of price continuation increases with capitalization for limit-up days ($N^{+}_{\rm{con}}/N^+$) and decreases with capitalization for limit-down days ($N^-_{\rm{con}}/N^-$). These trends are absent for $N^{\pm}_{\rm{close}}$. For bearish period, we observe that $N^-_{\rm{con}}/N^-$, $N^+_{\rm{close,con}}/N^+_{\rm{close}}$ and $N^-_{\rm{close,con}}/N^-_{\rm{close}}$ all have a decreasing trend with respect to the capitalization.

These findings have potential applications for practitioners in the Chinese stock markets, keeping in mind that one cannot short and there is a $T+1$ trading rule. When the price of a stock hits the up-limit, no matter in the intraday or at the close, she can buy right before market closure at 15:00 and sell with the opening price on the next trading day. Regardless of the transaction cost, the return will have high probability to be positive. The probability of earning money is even high if the price is locked to the up-limit and when the capitalization of the stock is low. Certainly, in this case, the liquidity is usually very low and it is hard to buy shares. On the contrary, if one holds a stock whose price experiences intraday down-limit hits, it is better to sell it to reduce losses.

\subsection*{Advanced statistics of intraday limit hits}

For each stock $i$ traded in $T_i$ days in our sample, we identify $K_i$ trading days on which the prices have hit either the up-limit or the down-limit at least once. We denote the set of limit-up days (trading days that have up-limit hits) of stock $i$ as ${\bf{U}}_i=\{u_{i,k}: k=1,2,\cdots,K^u_i\}$, where $K^u_i=\#{\bf{U}}_i$ is the number of limit-up trading days. Similarly, the set of limit-down days of stock $i$ is denoted as ${\bf{D}}_i=\{d_{i,k}: k=1,2,\cdots,K^d_i\}$, where $K^d_i=\#{\bf{D}}_i$ is the number of limit-down trading days. The intersection of ${\bf{U}}_i$ and ${\bf{D}}_i$ is not necessary to be empty, because, although very rare, it is possible that stock $i$ hits its up-limit and down-limit on the same day. In other words, $K_i\leq K_i^u+K_i^d$. The percents of limit-hitting days are calculated as follows
\begin{equation}
  \tilde{n}_i=\frac{{K_i}}{{T_i}},~~\tilde{n}_i^u=\frac{{K_i^u}}{T_i},~~{\mathrm{and}}~~\tilde{n}_i^d=\frac{{K_i^d}}{T_i},
  \label{Eq:Prob:Hit:per:Day}
\end{equation}
where $\tilde{n}_i$, $\tilde{n}_i^u$ and $\tilde{n}_i^d$ can be regarded as the empirical probabilities that stock $i$ will hit either the upper or the down-limit, the up-limit, and the down-limit on a trading day, respectively.

\begin{figure}[htb]
  \centering
  \includegraphics[width=0.96\textwidth,height=0.25\textwidth]{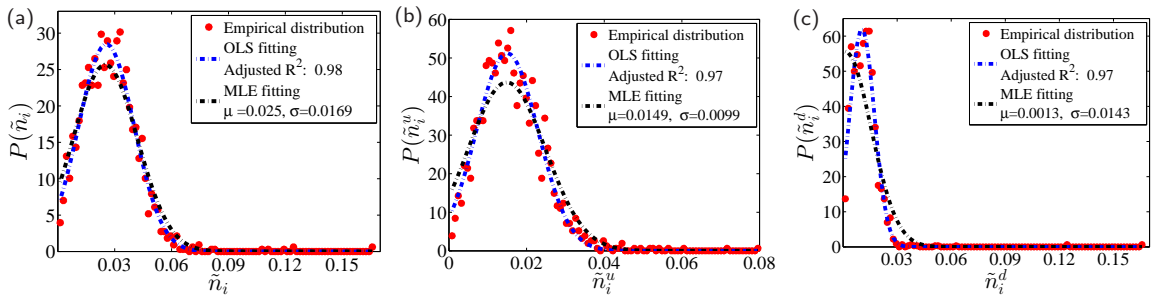}
  \caption{{\textbf{Empirical probability density functions $P(\tilde{n}_{i})$, $P(\tilde{n}^{u}_{i})$ and $P(\tilde{n}^{d}_{i})$ of daily price limit hitting probabilities $\tilde{n}_{i}$, $\tilde{n}^{u}_{i}$ and $\tilde{n}^{d}_{i}$ for individual stocks.}} The dots are empirical data, the black dash-dotted curves are the MLE fits to the truncated normal distribution in Eq.~(\ref{Eq:Normal:Left}), and the black dash-dotted curves are the OLS fits to the truncated normal distribution. (a) All limit hits; (b) Price up limit hits; (c) Price down limit hits.}
  \label{Fig:MagnetEffect:Prob:Hit:per:Day}
\end{figure}

Figure~\ref{Fig:MagnetEffect:Prob:Hit:per:Day} shows the empirical distributions of limit-hitting probability per day for individual stocks. As shown in Fig.~\ref{Fig:MagnetEffect:Prob:Hit:per:Day}(a), for most of the stocks, the daily limit-hitting probability is less than 6\%, with the most probability value around 2.5\%. However, there are also a few stocks having very large limit-hitting probability up to 17\%. Such seemingly outliers include 002606, 002632, 002635, 002643, 002644, and 002646. These stocks have relatively large numbers of price down limit hits, as shown in Fig.~\ref{Fig:MagnetEffect:Prob:Hit:per:Day}(c). The large values of $\tilde{n}_i^d$ are mainly caused by the small values of $T_i$ because these stocks come into the market for a short time period. For most stocks, the daily up-limit hit is less than 4\% according to Fig.~\ref{Fig:MagnetEffect:Prob:Hit:per:Day}(b) and the daily dow-limit hit is less than 3\% according to Fig.~\ref{Fig:MagnetEffect:Prob:Hit:per:Day}(c). This finding is consistent with the observation that $N^+>N^-$ in Table \ref{Tb:Statistics:ManyStocks}. It seems that the curves obtained by the OLS method fit the empirical data better than those by the MLE method.

Consider a limit-hitting day $u_{i,k}$ (or $d_{i,k}$) of stock $i$. The price may hit the price limit for $M^{u}_{i,k}$ (or $M^{d}_{i,k}$ ) times at intraday moments $t^{u}_{i,k,m}$ (or $t^{d}_{i,k,m}$) with $m=1, \cdots, M^{u}_{i,k}$ (or $M^{d}_{i,k}$). We define the average number of limit hits of stock $i$ as follows,
\begin{equation}
  M^{u}_i = \frac{{1}}{K^u_i}\sum_{k=1}^{K^u_i} M^u_{i,k}
  ~~{\mathrm{and}}~~
  M^{d}_i = \frac{{1}}{K^d_i}\sum_{k=1}^{K^d_i} M^{d}_{i,k}.
  \label{Eq:M:ud:i}
\end{equation}
where the denominators are not the total number of trading days, $T_i$, but the number of price up limit hitting days and of price down limit hitting days.

\begin{figure}[htb]
  \centering
  \includegraphics[width=0.66\textwidth,height=0.52\textwidth]{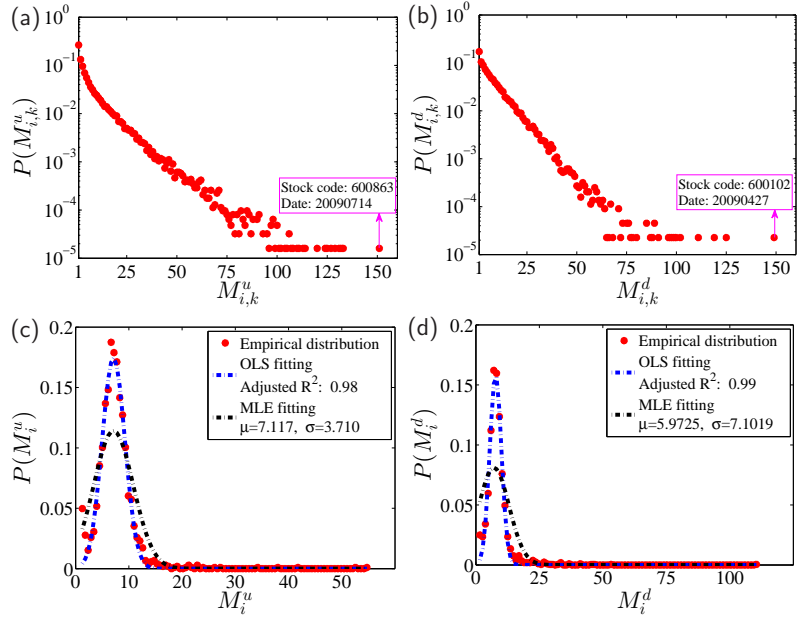}
  \caption{{\textbf{Number of limit hits on individual limit-hitting days.}} (a,b) Probability density functions $P(M^{u}_{i,k})$ and $P(M^{d}_{i,k})$ of $M^{u}_{i,k}$ and $M^{d}_{i,k}$. (c,d) Probability density functions $P(M^{u}_{i})$ and $P(M^{d}_{i})$ of $M^{u}_i$ and $M^{d}_i$ for individual stocks. }
  \label{Fig:MagnetEffect:PDF:Num:Hits}
\end{figure}

According to Fig.~\ref{Fig:MagnetEffect:PDF:Num:Hits}(a) and Fig.~\ref{Fig:MagnetEffect:PDF:Num:Hits}(b), the distributions of $M^{u}_{i,k}$ and $M^{d}_{i,k}$ decrease sharply. About 30\% of limit-hitting days have only one up-limit hit or down-limit hit. However, there are also limit-hitting days with very large numbers of limit hits in one day. For instance, stock 600863 hit the up-limit for 151 times on 2009/07/14, while stock 600102 hit the down-limit for 149 times on 2009/04/27, which has been delisted from the Shanghai Stock Exchange due to it poor performance. Surprisingly, the two distributions $P(M^{u}_{i})$ and $P(M^{d}_{i})$ are not monotonically decreasing functions. Instead, they can be well fitted by the truncated normal distribution. Speaking differently, if the price limit (10\% or -10\%) of a stock has been reached on certain trading day, the number of limit hits on that day is usually more than once.

When stock $i$ hits its price limit on day $u_{i,k}$ (or $d_{i,k}$), the $m$-th limit-up (or limit-down) started at $t^{u}_{i,k,m}$ (or $t^{d}_{i,k,m}$) may be opened at $t^{u}_{i,k,m}+\Delta{t}^{u}_{i,k,m}$ (or $t^{d}_{i,k,m}+\Delta{t}^{d}_{i,k,m}$). Hence, we have a sequence of limit-hitting durations $\Delta{t}^{u}_{i,k,m}$ (or $\Delta{t}^{d}_{i,k,m}$). The total duration on that day can be calculated as follows:
\begin{equation}
  \Delta{t}^{u}_{i,k} = \sum_{m=1}^{M^{u}_{i,k}} \Delta{t}^{u}_{i,k,m}
  ~~{\mathrm{and}}~~
  \Delta{t}^{d}_{i,k} = \sum_{m=1}^{M^{d}_{i,k}} \Delta{t}^{d}_{i,k,m}.
  \label{Eq:Delta:ud:i:k}
\end{equation}
Considering only the limit-hitting days of stock $i$, we can also define the average limit hit duration as follows:
\begin{equation}
  \Delta{t}^{u}_i = \frac{1}{K^u_i} \sum_{k=1}^{K^u_i} \sum_{m=1}^{M^{u}_{i,k}} \Delta{t}^{u}_{i,k,m}
  ~~{\mathrm{and}}~~
  \Delta{t}^{d}_i = \frac{1}{K^d_i}\sum_{k=1}^{K^d_i} \sum_{m=1}^{M^{d}_{i,k}} \Delta{t}^{d}_{i,k,m}.
  \label{Eq:Delta:ud:i}
\end{equation}
We stress that these two quantities are defined for individual stocks and the denominators are not $T_i$.

\begin{figure}[htb]
  \centering
  \includegraphics[width=0.96\textwidth,height=0.52\textwidth]{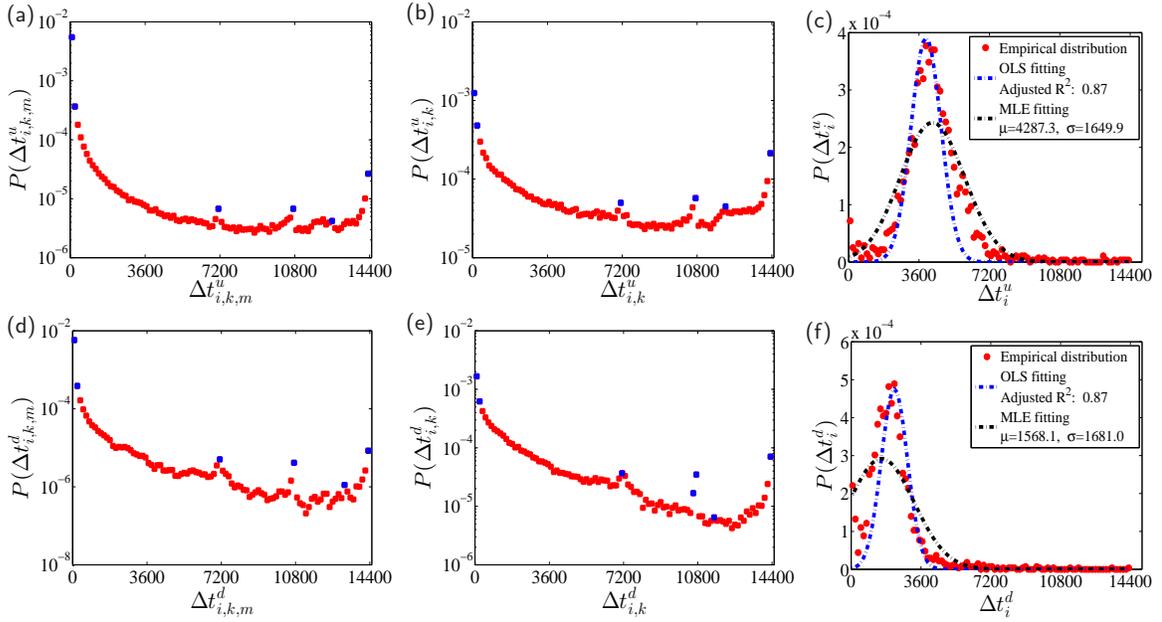}
  \caption{{\textbf{Limit hit duration.}} (a,d) Probability density functions $P(\Delta{t}^{u}_{i,k,m})$ and $P(\Delta{t}^{d}_{i,k,m})$ of the individual limit hit durations $\Delta{t}^{u}_{i,k,m}$ and $\Delta{t}^{d}_{i,k,m}$. (b,e) Probability density functions $P(\Delta{t}^{u}_{i,k})$ and $P(\Delta{t}^{d}_{i,k})$ of the daily limit hit durations $\Delta{t}^{u}_{i,k}$ and $\Delta{t}^{d}_{i,k}$. (c,f) Probability density functions $P(\Delta{t}^{u}_{i})$ and $P(\Delta{t}^{d}_{i})$  of the average daily limit hit durations $\Delta{t}^{u}_i$ and $\Delta{t}^{d}_i$. The unit of the variables is second.}
  \label{Fig:MagnetEffect:PDF:Duration}
\end{figure}

Plots (a) and (d) of Fig.~\ref{Fig:MagnetEffect:PDF:Duration} show the empirical distributions of the durations $\Delta{t}^{u}_{i,k,m}$ and $\Delta{t}^{d}_{i,k,m}$ of individual limit hits for all stocks. The two distributions exhibit a similar ``L'' shape. In each distribution, there are three obvious peaks at $\Delta{t}^{u,d}_{i,k,m}=7200$s, 10800s and 12600s, which correspond respectively to 2 hours, 3 hours and 3.5 hours. These peaks are mainly caused by limit hits at 13:00 p.m., 10:30 a.m. and 10:00 a.m. with the prices remaining at the limit prices till market closure. These patterns contain significant information contents. The peak around 13:00 reflects the fact that the majority of the traders or some informed traders may obtain cumulated important information about the holding stock during the market closure at noon (11:30-13:00) and form a collective behavior to buy or sell their stock to push the price to its limit. This interpretation applies certainly to the peak at 14400s, corresponding to the case that the stock opens at its limit price due to overnight information and does not fluctuate during the whole trading day. The peak at 10800s are caused by the one-hour opening trading halts in which the stock usually bears abnormal information pit and traders' trading habits and round number preference. The peak at 12600s is less significant, which is very likely caused by traders' trading habits and round number preference.

Plots (b) and (e) of Fig.~\ref{Fig:MagnetEffect:PDF:Duration} show the empirical distributions of the daily limit hit durations $\Delta{t}^{u}_{i,k}$ and $\Delta{t}^{d}_{i,k}$. The overall shapes of these two distributions are very similar with those in Fig.~\ref{Fig:MagnetEffect:PDF:Duration} (a) and (d). The main differences are that the distributions of daily limit hit durations have lower heights at the two edges and larger values in the bold parts. Plots (c) and (f) of Fig.~\ref{Fig:MagnetEffect:PDF:Duration} present the empirical distributions of the average daily durations $\Delta{t}^{u}_i$ and $\Delta{t}^{d}_i$ for individual stocks. The distributions are bimodal with an extra peak close to $\Delta{t}^{u,d}_i=0$s.

We further define the total intraday limit hit duration on trading day $u_k$, which is the time elapse from the moment of the first limit hit to the last moment that prices stay at limit in a limit-hitting day:
\begin{equation}
  \Delta{T}^{u}_{i,k} = t^{u}_{i,k,M^u_{i,k}}+\Delta{t}^{u}_{i,k,M^u_{i,k}}-t^{u}_{i,k,1}~~{\mathrm{and}}~~ \Delta{T}^{d}_{i,k} = t^{d}_{i,k,M^d_{i,k}}+\Delta{t}^{d}_{i,k,M^d_{i,k}}-t^{d}_{i,k,1}.
  \label{Eq:totalDelta:ud:i:k}
\end{equation}
By definition, we have $\Delta{T}^{u,d}_{i,k}\geq \Delta{t}^{u,d}_{i,k}$. We also define the average of the total intraday limit-hitting duration for individual stocks as follows,
\begin{equation}
  \Delta{T}^{u}_i = \frac{1}{K^u_i}\sum_{k=1}^{K^u_i}\Delta{T}^{u}_{i,k}~~{\mathrm{and}}~~  \Delta{T}^{d}_i = \frac{1}{K_i^d}\sum_{k=1}^{K^d_i} \Delta{T}^{d}_{i,k}.
  \label{Eq:totalDelta:ud:i}
\end{equation}
Similarly, we have $\Delta{T}^{u,d}_{i}\geq \Delta{t}^{u,d}_{i}$.

\begin{figure}[htb]
  \centering
  \includegraphics[width=0.66\textwidth,height=0.52\textwidth]{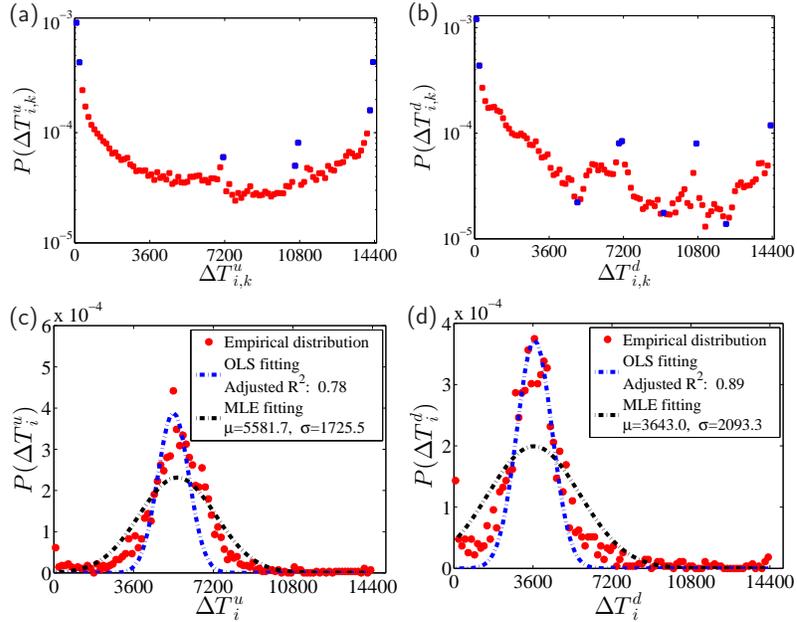}
  \caption{{\textbf{Total intraday limit hit duration.}} (a,b) Probability density functions $P(\Delta{T}^{u}_{i,k})$ and $P(\Delta{T}^{d}_{i,k})$ of the total intraday limit hit durations $\Delta{T}^{u}_{i,k}$ and $\Delta{T}^{d}_{i,k}$. (c,d) Probability density functions $P(\Delta{T}^{u}_{i})$ and $P(\Delta{T}^{d}_{i})$ of the averages of total intraday limit hit durations $\Delta{T}^{u}_i$ and $\Delta{T}^{d}_i$. The unit of the variables is second. }
  \label{Fig:MagnetEffect:PDF:totalDuration}
\end{figure}

The upper panel of Fig.~\ref{Fig:MagnetEffect:PDF:totalDuration} shows the empirical distributions of the total intraday limit hit durations $\Delta{T}^{u}_{i,k}$ and $\Delta{T}^{d}_{i,k}$ for all stocks. The distribution in Fig.~\ref{Fig:MagnetEffect:PDF:totalDuration}(a) has two local maxima around $\Delta{T}^{u}_{i,k}=7200$s and 10800s as in Fig.~\ref{Fig:MagnetEffect:PDF:Duration}(b) for $\Delta{t}^{u}_{i,k}$. The distribution in Fig.~\ref{Fig:MagnetEffect:PDF:totalDuration}(b) has a similar overall shape. However, it is less smoother. The lower panel of Fig.~\ref{Fig:MagnetEffect:PDF:totalDuration} shows the empirical distributions of $\Delta{T}^{u}_i$ and $\Delta{T}^{d}_i$. It is hard to find a suitable function form to fit the data.

As a complementary to the distributions in Fig.~\ref{Fig:MagnetEffect:PDF:Num:Hits} to Fig.~\ref{Fig:MagnetEffect:PDF:totalDuration}, we investigate the maxima, the medians and the means of the three kinds of daily variables, the daily numbers of limit hits $M^{u,d}_{i,k}$, the daily limit hit durations $\Delta{t}^{u,d}_{i,k}$ and the total intraday limit hit durations $\Delta{T}^{u,d}_{i,k}$. We compare the results for the whole time period, the bullish periods and the bearish periods. We also divide all the stocks into six portfolios as defined in Table \ref{Tb:Statistics:ManyStocks} to investigate the impacts of stock capitalization. The results are presented in Table \ref{Tb:Intraday:statistics:IndividualStocks}.

\begin{table}[!ht]
%  \begin{adjustwidth}{-2.25in}{0in} % Comment out/remove adjustwidth environment if table fits in text column.
  \caption{{\textbf{Statistics of intraday limit hits for individual stocks.}} The determination of the six portfolios is the same as in Table \ref{Tb:Statistics:ManyStocks}. The maxima, the mediates and the means of the daily numbers of limit hits $M^{u,d}_{i,k}$, the daily limit hit durations $\Delta{t}^{u,d}_{i,k}$ and the total intraday limit hit durations $\Delta{T}^{u,d}_{i,k}$ are presented. Results for bullish periods and bearish periods are compared.}
\begin{tabular}{cccccccccccccccccccccccccccccccccccccccccccccc}
\hline\hline
\multicolumn{8}{l}{{\small{Panel A: Whole sample period}}}\\\hline
  && \multicolumn{2}{c}{All stocks} && \multicolumn{2}{c}{Portfolio 1} && \multicolumn{2}{c}{Portfolio 2} && \multicolumn{2}{c}{Portfolio 3} && \multicolumn{2}{c}{Portfolio 4} && \multicolumn{2}{c}{Portfolio 5} && \multicolumn{2}{c}{Portfolio 6}\\
  \cline{3-4}\cline{6-7}\cline{9-10}\cline{12-13}\cline{15-16}\cline{18-19}\cline{21-22}
  &&Up & Down && Up & Down && Up & Down && Up & Down && Up & Down && Up & Down  && Up & Down  \\\hline
%${\min}\{M^{u,d}_{i,k}\}$               &&1 & 1 && 1 & 1 && 1 & 1 && 1 & 1 && 1 & 1 && 1 & 1 && 1 & 1\\
  ${\max}\{M^{u,d}_{i,k}\}$               && 151& 149&& 88& 87&& 114& 102&& 113& 96&& 125& 119&& 133& 99&& 151& 149\\
  ${\rm{mean}}\{M^{u,d}_{i,k}\}$          &&7.47 & 8.27 && 5.74 & 6.67 && 7.13 & 7.86 && 7.70 & 8.62 && 7.95 & 8.55 && 8.06 & 8.83 && 8.23 & 9.09\\
  ${\rm{med}}\{M^{u,d}_{i,k}\}$           && 4& 5&& 3& 4&& 4& 5&& 4& 6&& 4& 6&& 4& 6&& 4& 6\\
  %${\min}\{\Delta{t}^{u,d}_{i,k}\}$       && 0 & 0 && 0 & 0 && 0 & 0 && 0 & 0 && 0 & 0 && 0 & 0 && 0 & 0 \\
  %${\max}\{\Delta{t}^{u,d}_{i,k}\}$       &&14400& 14398&& 14396& 14397&& 14400& 14393&& 14398& 14395&& 14400& 14396&& 14398& 14391&& 14400& 14398\\
  ${\rm{mean}}\{\Delta{t}^{u,d}_{i,k}\}$  && 4391 & 2189 && 4285 & 2014 && 4723 & 2036 && 4696 & 2291 && 4542 & 2347 && 4300 & 2247 && 3801 & 2202\\
  ${\rm{med}}\{\Delta{t}^{u,d}_{i,k}\}$   &&2096& 815&& 1720& 552&& 2386& 745&& 2397& 906&& 2282& 911&& 2113& 866&& 1815& 917\\
  %${\min}\{\Delta{T}^{u,d}_{i}\}$                 &&0 & 0 && 0 & 0 && 0 & 0 && 0 & 0 && 0 & 0 && 0 & 0 && 0 & 0\\
  %${\max}\{\Delta{T}^{u,d}_{i}\}$                 && 14400& 14398&& 14399& 14397&& 14400& 14395&& 14398& 14395&& 14400& 14397&& 14398& 14397&& 14400& 14398\\
  ${\rm{mean}}\{\Delta{T}^{u,d}_{i}\}$            &&5740 & 3888 && 5553 & 3793 && 6241 & 3661 && 6059 & 4035 && 5822 & 4054 && 5590 & 3924 && 5175 & 3857\\
  ${\rm{med}}\{\Delta{T}^{u,d}_{i}\}$             &&4118& 1937&& 3496& 1785&& 5020& 1796&& 4740& 2035&& 4324& 2077&& 4004& 1953&& 3408& 1982\\
  \hline\multicolumn{8}{l}{Panel B: Bull period}\\\hline
  && \multicolumn{2}{c}{All stocks} && \multicolumn{2}{c}{Portfolio 1} && \multicolumn{2}{c}{Portfolio 2} && \multicolumn{2}{c}{Portfolio 3} && \multicolumn{2}{c}{Portfolio 4} && \multicolumn{2}{c}{Portfolio 5} && \multicolumn{2}{c}{Portfolio 6}\\
  \cline{3-4}\cline{6-7}\cline{9-10}\cline{12-13}\cline{15-16}\cline{18-19}\cline{21-22}
  &&Up & Down && Up & Down && Up & Down && Up & Down && Up & Down && Up & Down  && Up & Down  \\\hline
  %${\min}\{M^{u,d}_{i,k}\}$               &&1 & 1 && 1 & 1 && 1 & 1 && 1 & 1 && 1 & 1 && 1 & 1 && 1 & 1\\
  ${\max}\{M^{u,d}_{i,k}\}$               && 151& 149&& 88& 87&& 97& 102&& 114& 96&& 113& 72&& 125& 97&& 151& 149\\
  ${\rm{mean}}\{M^{u,d}_{i,k}\}$          &&7.34 & 7.76 && 6.42 & 7.16 && 7.19 & 7.67 && 7.71 & 8.18 && 7.64 & 7.83 && 7.46 & 7.90 && 7.61 & 7.83\\
  ${\rm{med}}\{M^{u,d}_{i,k}\}$           && 4& 5&& 3& 4&& 4& 5&& 4& 5&& 4& 5&& 4& 5&& 4& 5\\
  %${\min}\{\Delta{t}^{u,d}_{i,k}\}$       && 0 & 0 && 0 & 0 && 0 & 0 && 0 & 0 && 0 & 0 && 0 & 0 && 0 & 0 \\
  %${\max}\{\Delta{t}^{u,d}_{i,k}\}$       &&14400& 14398&& 14396& 14397&& 14400& 14393&& 14398& 14395&& 14400& 14396&& 14398& 14391&& 14400& 14398\\
  ${\rm{mean}}\{\Delta{t}^{u,d}_{i,k}\}$  && 4064 & 2403 && 3956 & 1757 && 4276 & 1991 && 4327 & 2643 && 4288 & 2858 && 4011 & 2719 && 3525 & 2452\\
  ${\rm{med}}\{\Delta{t}^{u,d}_{i,k}\}$   &&1778 & 790&& 1290 & 408&& 1901& 603&& 2070 & 956&& 2093 & 1064&& 1850& 1016&& 1580& 965\\
  %${\min}\{\Delta{T}^{u,d}_{i}\}$                 &&0 & 0 && 0 & 0 && 0 & 0 && 0 & 0 && 0 & 0 && 0 & 0 && 0 & 0\\
  %${\max}\{\Delta{T}^{u,d}_{i}\}$                 && 14400& 14398&& 14399& 14397&& 14400& 14395&& 14398& 14395&& 14400& 14397&& 14398& 14397&& 14400& 14398\\
  ${\rm{mean}}\{\Delta{T}^{u,d}_{i}\}$            &&5251 & 4011 && 5123 & 3566 && 5410 & 3521 && 5524 & 4307 && 5493 & 4388 && 5225 & 4286 && 4736 & 3997\\
  ${\rm{med}}\{\Delta{T}^{u,d}_{i}\}$             &&3415 & 2018&& 2689& 1539&& 3675& 1605&& 3982 & 2187&& 3877 & 2264&& 3495& 2213&& 2853 & 2089\\
  \hline\multicolumn{8}{l}{Panel C: Bear period}\\\hline
  && \multicolumn{2}{c}{All stocks} && \multicolumn{2}{c}{Portfolio 1} && \multicolumn{2}{c}{Portfolio 2} && \multicolumn{2}{c}{Portfolio 3} && \multicolumn{2}{c}{Portfolio 4} && \multicolumn{2}{c}{Portfolio 5} && \multicolumn{2}{c}{Portfolio 6}\\
  \cline{3-4}\cline{6-7}\cline{9-10}\cline{12-13}\cline{15-16}\cline{18-19}\cline{21-22}
  &&Up & Down && Up & Down && Up & Down && Up & Down && Up & Down && Up & Down  && Up & Down  \\\hline
    %${\min}\{M^{u,d}_{i,k}\}$               &&1 & 1 && 1 & 1 && 1 & 1 && 1 & 1 && 1 & 1 && 1 & 1 && 1 & 1\\
  ${\max}\{M^{u,d}_{i,k}\}$               && 133& 125&& 65& 59&& 88& 71&& 108& 73&& 133& 119&& 131& 99&& 124& 125\\
  ${\rm{mean}}\{M^{u,d}_{i,k}\}$          &&7.61 & 8.62 && 4.84 & 6.30 && 7.23 & 8.02 && 7.96 & 8.99 && 8.28 & 9.05 && 8.55 & 9.58 && 8.81 & 9.78\\
  ${\rm{med}}\{M^{u,d}_{i,k}\}$           && 3& 6&& 2& 4&& 4& 6&& 4& 6&& 4& 6&& 4& 7&& 4& 7\\
  %${\min}\{\Delta{t}^{u,d}_{i,k}\}$       && 0 & 0 && 0 & 0 && 0 & 0 && 0 & 0 && 0 & 0 && 0 & 0 && 0 & 0 \\
  %${\max}\{\Delta{t}^{u,d}_{i,k}\}$       &&14400& 14399&& 14400& 14393&& 14398& 14397&& 14398& 14395&& 14399& 14396&& 14399& 14397&& 14399& 14399\\
  ${\rm{mean}}\{\Delta{t}^{u,d}_{i,k}\}$  && 4750 & 2044 && 4912 & 2193 && 5227 & 2042 && 5065 & 2013 && 4764 & 1988 && 4468 & 1979 && 4064 & 2049\\
  ${\rm{med}}\{\Delta{t}^{u,d}_{i,k}\}$   &&2520& 829&& 2848& 707&& 3093& 878&& 2640& 855&& 2541& 817&& 2305& 819&& 2043& 885\\
  %${\min}\{\Delta{T}^{u,d}_{i}\}$                 &&0 & 0 && 0 & 0 && 0 & 0 && 0 & 0 && 0 & 0 && 0 & 0 && 0 & 0\\
  %${\max}\{\Delta{T}^{u,d}_{i}\}$                 && 14400& 14399&& 14400& 14399&& 14398& 14397&& 14399& 14397&& 14399& 14396&& 14399& 14397&& 14399& 14399\\
  ${\rm{mean}}\{\Delta{T}^{u,d}_{i}\}$            &&6275 & 3804 && 6460 & 3950 && 7179 & 3717 && 6470 & 3830 && 6104 & 3825 && 5838 & 3753 && 5597 & 3747\\
  ${\rm{med}}\{\Delta{T}^{u,d}_{i}\}$             &&5030& 1871&& 5509& 1919&& 7172& 1917&& 5480& 1869 && 4905& 1814&& 4305& 1845&& 4002 & 1885\\
  \hline\hline
  \end{tabular}
  %\begin{flushleft} Table notes Phasellus venenatis, tortor nec vestibulum mattis, massa tortor interdum felis, nec pellentesque metus tortor nec nisl. Ut ornare mauris tellus, vel dapibus arcu suscipit sed.
%  \end{flushleft}
  \label{Tb:Intraday:statistics:IndividualStocks}
%  \end{adjustwidth}
\end{table}

For the maximal number of limit hits, we find that ${\max}\{M^{u}_{i,k}\}>{\max}\{M^{d}_{i,k}\}$ for all cases, except for Portfolio 2 in bullish periods. However, for the means, we have ${\rm{mean}}\{M^{u}_{i,k}\}<{\rm{mean}}\{M^{d}_{i,k}\}$ in all the cases. On average, down-limit hits are less stable than up-limit hits, indicating that the fear sentiment of traders fluctuates more than the greed sentiments in the Chinese stock markets. Similarly, we observe that  ${\rm{med}}\{M^{u}_{i,k}\}<{\rm{med}}\{M^{d}_{i,k}\}$ in all the cases. However, the difference ${\rm{med}}\{M^{d}_{i,k}\}-{\rm{med}}\{M^{u}_{i,k}\}$ is larger in the bearish periods than in the bullish periods. It indicates that the fighting between long positions and short positions is more severe when the price hits the down-limit than the up-limit, because trades can make profits when the market rises up while they can only deduce losses when the market falls down. No clear correlation is found between ${\rm{mean}}\{M^{u,d}_{i,k}\}$ and capitalization in the bullish periods. In contrast, both ${\rm{mean}}\{M^{u}_{i,k}\}$ and ${\rm{mean}}\{M^{d}_{i,k}\}$ are positively correlated with capitalization in the bearish periods.

For the other variables, we find that the quantities for up-limits are greater than their counterparts for down-limits. Specifically, we observe that ${\rm{mean}}\{\Delta{t}^{u}_{i,k}\}>{\rm{mean}}\{\Delta{t}^{d}_{i,k}\}$, ${\rm{med}}\{\Delta{t}^{u}_{i,k}\}>{\rm{med}}\{\Delta{t}^{d}_{i,k}\}$, ${\rm{mean}}\{\Delta{T}^{u}_{i,k}\}>{\rm{mean}}\{\Delta{T}^{d}_{i,k}\}$, and ${\rm{med}}\{\Delta{T}^{u}_{i,k}\}>{\rm{med}}\{\Delta{T}^{d}_{i,k}\}$ for different periods and different portfolios. Together with the results for ${\rm{mean}}\{M^{u,d}_{i,k}\}$ , we find that the average duration of individual up-limit hits is about twice longer than that of individual down-limit hits. This is probably caused by the no-short trading rule because traders can make money only when the market rises. There are also traders known as ``dare-to-die corps for up-limit hits'', who are actively engaged in pushing prices to the up-limits. It is also found that these quantities have a bell-like shape or monotonically decrease with respect to the capitalization.

\subsection*{Intraday patterns of the occurrence of price limit hits}

We divide the continuous double auction period in each trading day into $N$ intervals of equal length of $\Delta{t}$ minutes. If an interval starts at $t_0$ and ends at $t_1=t_0+\Delta{t}$, we can count the number of occurrences of up-limit hits in this interval as
\begin{equation}
  C^u_i = \sum_{k=1}^{K^u_i}{\rm{I}}_{t^u_{i,k}\in(t_0,t_1]},
  \label{Eq:C:u:i}
\end{equation}
and similarly the number of occurrences of down-limit hits
\begin{equation}
  C^d_i = \sum_{k=1}^{K^d_i}{\rm{I}}_{t^d_{i,k}\in(t_0,t_1]},
  \label{Eq:C:d:i}
\end{equation}
where ${\rm{I}}_x$ is an indicator function of event such that the value of ${\rm{I}}_x$ is 1 if the event $x$ is true and 0 otherwise. We can further calculate the following quantities
\begin{equation}
  C^{u,d} = \sum_{i=1}^I{C^{u,d}_i},
  \label{Eq:C:u:d}
\end{equation}
where $I$ is the number of stocks in the investigated sample.

\begin{figure}[!htb]
  \centering
  \includegraphics[width=0.96\textwidth,height=0.52\textwidth]{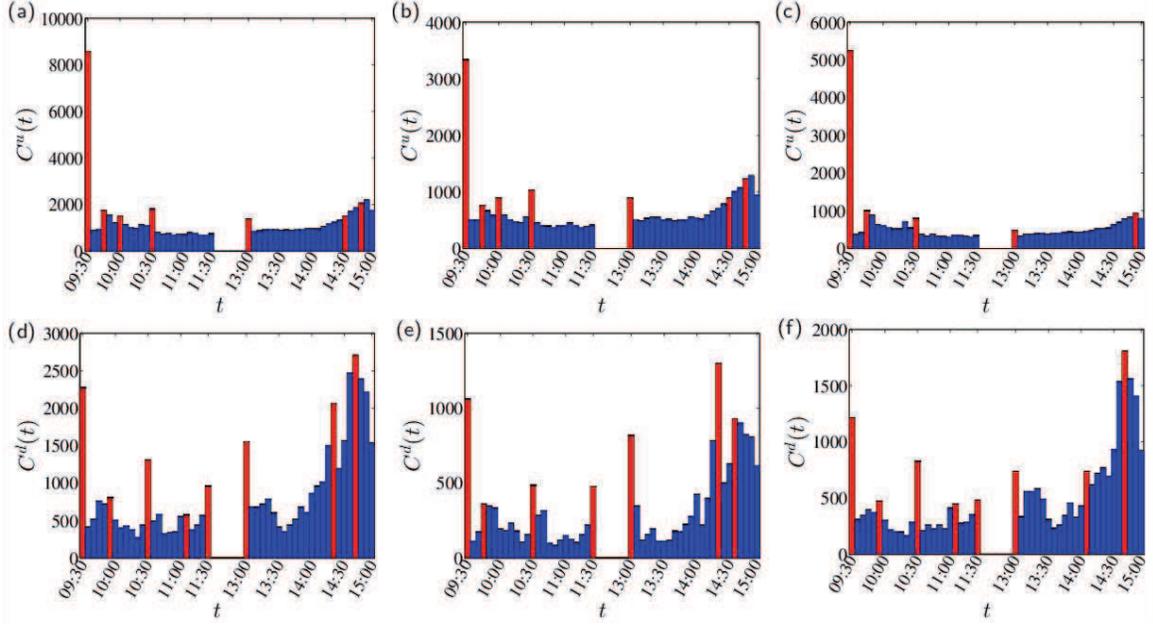}
  \caption{{\textbf{Intraday patterns of the occurrence of limit hits.}} The three columns of plots correspond respectively to the whole period (a,d), the bullish periods (b,e) and the bearish periods (c,f). The time interval $\Delta{t}$ is 5 min.}
  \label{Fig:Interval:C:ud}
\end{figure}

The intraday patterns of the occurrence of limit hits in intervals of size $\Delta{t}=5$ min are illustrated in Fig.~\ref{Fig:Interval:C:ud}. For each intraday interval, we have $C^{u,d} = C^{u,d}_{\rm{bull}} + C^{u,d}_{\rm{bear}}$. The occurrence of up-limit hits is extremely high at the opening of the market around 9:30 a.m., while the occurrence of down-limits is relatively high at the opening. There are also other local peaks around 09:30, 09:45, 10:00, 10:30, 11:30, 13:30, 14:30 and 14:45. Up-limit hits have different intraday patterns from down-limit hits. However, the intraday patterns do not differ much when comparing the bullish periods with the bearish periods. In all the six cases, limit hits are more frequent in the last hour, especially for down-limit hits. For up-limits, the occurrence number is quite stable over the period from 10:30 a.m. to 14:20 p.m., except for the afternoon opening of the market.

\subsection*{Dynamics of financial variables before limit hits}

Limit hits are rare events. The dynamics of financial variables before limits can enrich our understanding of trading activities of investors around such extreme events. We investigate here the evolution of several important financial variables before stock prices hit the up-limits and down-limits. Limit hits occurred at the opening of the market are excluded from analysis, because one cannot trace the pre-event dynamics of any financial variables.

We first study the average velocity of price change for all limit-hitting events in isometric intervals of price changes during bullish periods and bearish periods. For each limit-hitting event $i$, we consider the cases when the price rises above 5\% or falls below -5\%, compared with the close price of the previous trading day. We divide each of the two intervals $(5\%,10\%]$ and $[-10\%,-5\%)$ into 10 subintervals $((5 + 0.5m)\% ,(5.5 + 0.5m)\%]$ and $[-(5.5 + 0.5m)\%,-(5 + 0.5m)\%)$, where $m = 0,1,2,\cdots,9$. Let $\Delta{t}_{im}$ denotes the time duration for the stock price rising from $(5 + 0.5m)\%$ to $(5.5 + 0.5m)\%]$ or dropping from $-(5 + 0.5m)\%$ to $-(5.5 + 0.5m)\%]$. We consider four classes of limit hits: up-limit hits in bullish periods, up-limit hits in bearish periods, down-limit hits in bullish periods, and down-limit hits in bearish periods. For each class of limit hits, we define the dimensionless velocity of price change as follow,
\begin{equation}
\label{Eq:period:yield:velocity}
  V_{m} = \frac{1}{{\frac{1}{\mathcal{N}}\sum\limits_{i = 1}^{\mathcal{N}} {\frac{{\Delta {t_{im}}}}{{\sum\limits_{m = 0}^9 {\Delta {t_{im}}} }}} }}, ~~ ~~ m = 0,1,2,\cdots,9\
\end{equation}
where $\mathcal{N}$ is the number of limit hits of the class under investigation. If $\Delta{t}_{im}$ is independent of $m$ for all $i$, the velocity is constant such that $V_m=9$. If the average $\langle\Delta{t}_{im}\rangle$ over $i$ increases (decreases) with $m$, $V_{m}$ decreases (increases) when the price approaches the price limit, showing evidence of a cooling off (magnet) effect.

\begin{figure}[htb]
  \centering
  \includegraphics[width=0.96\textwidth,height=0.36\textwidth]{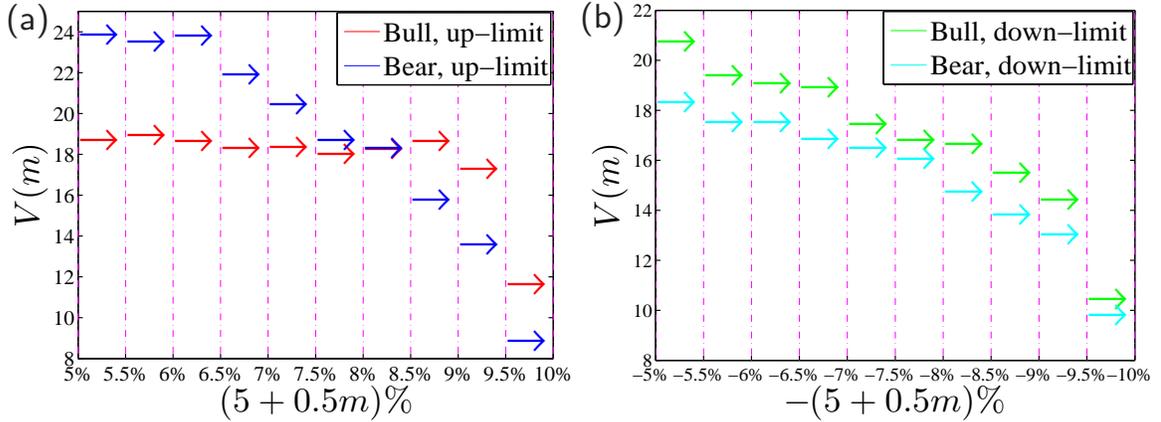}
  \caption{\label{Fig:Prehit:Price:Acceleration} {\textbf{Cooling-off effect of the price limit rule.}} (a) Comparison of price movement velocity $V_m^u$ before up-limit hits in the bullish and bearish periods. (b) Comparison of price movement velocity $V_m^d$ before down-limit hits in the bullish and bearish periods.}
\end{figure}

As shown in Fig.~\ref{Fig:Prehit:Price:Acceleration}, when stock price approaches the limit price, the price movement velocity decreases, except that the velocity does not change much when the price rise is less than 9\% for up-limit hits in the bullish periods. It suggests that there exists cooling-off effects when stock price gets close to the limit price. In this sense, the price limit rule is effective in the Chinese stock markets. The cooling-off effect is more remarkable in bearish periods than in bullish periods before up-limit hits and down-limit hits. However, the effectiveness of the cooling-off effect is mixed between up-limit hits and down-limit hits. The asymmetry of the cooling-off effect between bullish and bearish periods is probably caused by the no-short trading rule and the special lift forces in the bullish periods, such as the ``dare-to-die corps for up-limit hits''.

We now study the evolution of another four financial measures associated with the last 100 trades till hitting the price limits. We first investigate the evolution of sizes of buyer-initiated trades and seller-initiated trades on the last 100 trades including the one pushing the price to its limit. The last trade causes the price to hit its limit, labelled as the 100th trade. If the $k$-th trade was initiated by a buyer before an up-limit hit $i$ or by a seller before a down-limit hit $i$, we denote $s_i^+(k)$ as its size and, in this case, $s_i^-(k)=0$. Alternatively, if the $k$-th trade was initiated by a seller before an up-limit hit $i$ or by a buyer before a down-limit hit $i$, we denote $s_i^-(k)$ as its size and $s_i^+(k)=0$. Hence, $s_i^+(k)$ and $s_i^-(k)$ are respectively the sizes of same-direction (momentum) and opposite-direction (contrarian) trades that move the price towards and away from the limit price. We compare up-limit hits and down-limit hits. We also consider separately bullish periods and bearish periods. The average logarithmic trade sizes of the last 100 transactions are calculated as follows,
\begin{equation}
  s^{\pm}(k) = \frac{1}{\mathcal{N}}\sum\limits_{i = 1}^{\mathcal{N}} \ln\left[s_i^{\pm}(k)\right],
  \label{Eq:Prehit:TradeSizes}
\end{equation}
where $\mathcal{N}$ is the number of limit hits in one of the four cases (up-limit in bullish periods, down-limit in bullish periods, up-limit in bearish periods, and down-limit in bearish periods). Since the last trade is by definition the same-direction trade that push the price to the limit, we have $s^-_i(100)\equiv 0$ in all cases. For simplicity, we use the notation $s^-(100)=0$ instead of $s^-(100)=-\infty$ in the following.

\begin{figure}[htb]
  \centering
  \includegraphics[width=0.8\textwidth,height=0.6\textwidth]{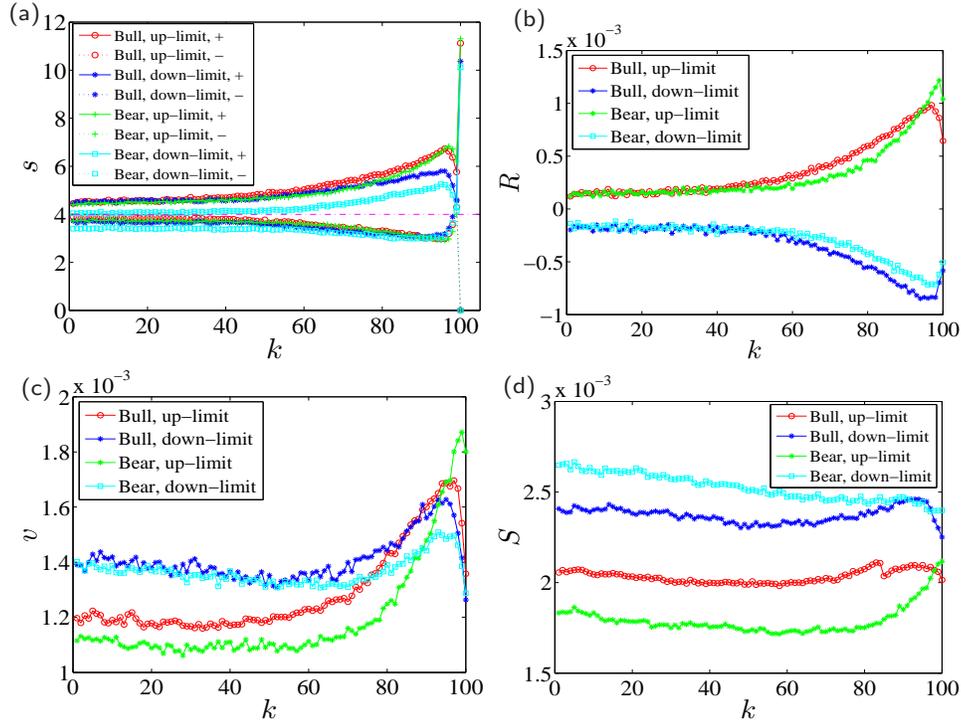}
  \caption{\label{Fig:Prehit:100trades:4quantities} {\textbf{Evolution of four important financial quantities along the last 100 trades right before limit hits.}} (a) Evolution of the logarithmic same-direction trade sizes $s^+(k)$ (the upper bundle) and the logarithmic opposite-direction trade sizes $s^-(k)$ (the lower bundle) for up-limit hits and down-limit hits in the bullish periods and bearish periods. (b) Evolution of the average trade-by-trade return $R(k)$. (c) Evolution of the average trade-by-trade volatility $v(K)$. (d) Evolution of the average bid-ask spread right before individual trades. For each financial quantity, we consider four cases, that is, up-limit hits in bullish periods, down-limit hits in bullish periods, up-limit hits in bearish periods, and down-limit hits in bearish periods.}
\end{figure}

Figure \ref{Fig:Prehit:100trades:4quantities}(a) illustrates the evolution of the same-direction trade sizes $s^+(k)$ and the opposite-direction trade sizes $s^-(k)$ for different cases. The same-direction trade sizes $s^+(k)$ have similar evolutionary trajectories in all the four cases, so do the opposite-direction trade sizes $s^-(k)$. We find that $s^+(k)$ increases from $k=1$ to reach the local maximum at $k=96$ and then decreases till $k=99$ followed by a very large $s^+(100)$ pushing the price to the limit, while $s^-(k)$ decreases from $k=1$ to reach the local minimum at $k=96$ and then increases till $k=99$, ended with $s^-(100)=0$. For all $k$, we have $s^+(k)>s^-(k)$, indicating that the pushing force of the same-direction traders is stronger than that of the opposite-direction traders. The trade size difference $s^+(k)-s^-(k)$ continuously increases and reaches its local maxima at $k=96$ and then decreases till $k=99$. These findings also provide clues about the cooling down of the last three trades right before limit hits. The trade sizes $s^+(k)$ are larger when the price is close to the up-limit than to the down-limit, which is true for both bullish and bearish periods. The trade sizes $s^+(k)$ for the case of down-limit hits in bearish periods are relatively the lowest. The relative position of the four $s^+(k)$ curves provides evidence of a well-known trait of Chinese traders that they tend to buy rising stocks and hate to sell holding shares when the stock price drops.

As shown in Fig.~\ref{Fig:Prehit:100trades:4quantities}(a), the average size $s^+(100)$ of the last trade that pushed the price to hit the limit is extraordinary larger than the size of any preceding trades. We show the empirical distributions of $s^+(100)$ for the four cases. All the distributions are unimodal and have similar shapes. However, they have different peak heights. The peak is the highest for the case of up-limit hits in the bullish periods, the second largest for the case of up-limit hits in the bearish periods.

We denote $p_i(k)$ the transaction price of the $k$-th trade before the $i$-th limit hit, $\left\{p^a_{i,j}(k)|j=1,\cdots,J\right\}$ and $\left\{V^a_{i,j}(k)|j=1,\cdots,J\right\}$ the prices and standing volumes at the first $J$ price levels of the sell-side limit order book right before the $k$-th trade, and $\left\{p^b_{i,j}(k)|j=1,\cdots,J\right\}$ and $\left\{V^b_{i,j}(k)|j=1,\cdots,J\right\}$ the prices and standing volumes on the first $J$ price levels of the buy-side limit order book right before the $k$-th trade. The second financial variable investigated is the average trade-by-trade return which is defined as follows,
\begin{equation}
  R(k) = \frac{1}{\mathcal{N}}\sum\limits_{i = 1}^{\mathcal{N}} \ln p_{i}(k) - \ln p_{i}(k-1), ~~ k= 1,2,3,\cdots,100.
  \label{Eq:Prehit:100trades:Return}
\end{equation}
Figure \ref{Fig:Prehit:100trades:4quantities}(b) shows the evolution of average trade-by-trade return $R(k)$ before price limit hits. It is found that $R(k)$ increases superlinearly before $k$ close to 100 and then decreases sharply before up-limit hits in both bullish and bearish periods. Almost symmetrically, $R(k)$ decreases superlinearly before $k$ close to 100 and then increases sharply before down-limit hits in both bullish and bearish periods. These patterns are consistent with the behaviors of the average trade sizes, because large trade sizes usually cause large price movements \cite{Lillo-Farmer-Mantegna-2003-Nature,Lim-Coggins-2005-QF,Zhou-2012-QF,Zhou-2012-NJP}.

The third variable is the average trade-by-trade volatility which is defined as follows,
\begin{equation}
  v(k) = \frac{1}{\mathcal{N}}\sum\limits_{i = 1}^{\mathcal{N}} |\ln p_{i}(k) - \ln p_{i}(k-1)|, ~~ k= 1,2,3,\cdots,100.
  \label{Eq:Prehit:100trades:Volatility}
\end{equation}
Figure \ref{Fig:Prehit:100trades:4quantities}(c) presents the evolution of average trade-by-trade volatility $v(k)$ before price limit hits. The volatility $v(k)$ before up-limit hits for both bullish and bearish periods is relatively stable when $k$ is less than about 50, then increases rapidly to reach a maximum one or two trades before limit hits, and finally drops to some extent. About five trades before up-limit hits, the volatility is higher during bullish periods than in bearish periods. The volatility about 70 trades before down-limit hits is higher than that before up-limit hits and exhibits a mild decreasing trend. The volatility increases afterwards and decreases again around $k=95$.

The fourth variable investigated is the average bid-ask spread right before the $k$-th trade which is defined as follows,
\begin{equation}
  S(k) = \frac{1}{\mathcal{N}}\sum\limits_{i = 1}^{\mathcal{N}} \frac{p^a_{i,1}(k) - p^b_{i,1}(k)}{\frac{1}{2}\left[p^a_{i,1}(k) + p^b_{i,1}(k)\right]}, ~~ k= 1,2,3,\cdots,100.
  \label{Eq:Prehit:100trades:Spread}
\end{equation}
As illustrated in Fig.~\ref{Fig:Prehit:100trades:4quantities}(d), the four curves of the average bid-ask spread $S(k)$ decrease before $k\approx60$. The two spread curves for the bullish periods have very similar shapes. After the initial decrease, they increase during the next twenty trades or so and decrease again before the limit hits. For the curve associated with up-limit hits in bearish periods, the spread increases continuously after $k\approx75$. The comparison of the four curves is also quite intriguing. The spread before up-limit hits is narrower than that before down-limit hits, indicating higher liquidity before up-limit hits. On average, the spread is the narrowest before up-limit hits in bearish periods and the widest before down-limit hits in bearish periods.

\section*{Discussion}

Stock markets are complex systems in which humans interact by buying and selling shares. The evolutionary trajectories of stock markets are fully determined by the behaviors of human being. It is widely accepted that people in emerging markets are less skilled and more irrational and thus these markets are much riskier. Due to different factors such as imitation, global news, as well as collusive manipulation, traders may herd to push the price rise up or drop down rapidly in very short time intervals through positive feedback loops. Such kind of collective behaviors might be caused partially by the somewhat reciprocity among a small amount of traders through price manipulation, which is reminiscent of the cooperation phenomena among human beings \cite{Nowak-2005-Science,Perc-Szolnoki-2010-Bs,Perc-GomezGardenes-Szolnoki-Floria-Moreno-2013-JRSI,Szolnoki-Perc-2013-PRX}.

To cool off traders' intraday mania to avoid the price deviating much from its fundamental value, a number of stock exchanges pose price limit rules. Empirical evidence is controversial about the presence of a magnet effect or a cooling off effect even for the Chinese stock markets \cite{Zhang-Zhu-2014-cnJCQUT}. Our preliminary result on the price movement velocity favors the presence of an intraday cooling-off effect for both up-limits and down-limits in both bullish and bearish periods. The evolution of other financial variables such as trade size, trade-by-trade return, trade-by-trade volatility and bid-ask spread seems to support the conclusion that the traders are cooled off right before stock prices hit the price limits, as shown by the anti-trend behavior just before hitting the price limits. It is well documented that the limit order books are thin near price limits \cite{Gu-Chen-Zhou-2008c-PA} and thus the liquidity is worse. When the price is pushed towards its limit, traders submit larger and larger orders and the price changes enlarge. However, the population of traders who place opposite orders to realize their gains or act as bottom fishers also increases. When the price is close to its limit, the force of opposite traders reverses the trends. In this sense, the price limit rule works in the Chinese stock markets.

We have investigated the statistical properties of characteristic variables of up-limit and down-limit hits in bullish and bearish periods. We also uncovered nonlinear impacts of stock capitalization on price limit hits by comparing six portfolios sorted due to stocks' capitalization on the daily level. It is intriguing to find that price continuation occurs more frequently than price reversal on the next trading day after a limit-hitting day. This effect is more significant for down-limit hits. The empirical probability of next-day price continuation is thus far greater than 50\%. Our empirical findings have potential practical values for market practitioners. For instance, it will be probably profitable to buy shares at the close price in a up-limit hitting day and sell the share at the opening of the next trading day, or to sell the shares one holds in a down-limit hitting day.

\end{document}